\documentclass[12pt,preprint]{aastex}

\slugcomment{To appear in AJ, September 7, 2010}

\shorttitle{Kuiper Belt Dust}
\shortauthors{Kuchner and Stark}
\begin{document}

\title{Collisional Grooming Models of the Kuiper Belt \\ Dust Cloud}

\author{Marc J. Kuchner}
\affil{NASA Goddard Space Flight Center \\
Exoplanets and Stellar Astrophysics Laboratory, Code 667
\\ Greenbelt, MD 21230}
\email{Marc.Kuchner@nasa.gov}

\author{Christopher C. Stark}
\affil{Department of Physics, University of Maryland, Box 197 \\ 082 Regents Drive \\ College Park, MD 20742-4111}
\email{starkc@umd.edu}

\begin{abstract}

We modeled the 3-D structure of the Kuiper Belt dust cloud at four different dust production rates, incorporating both planet-dust interactions and grain-grain collisions using the collisional grooming algorithm.   Simulated images of a model with a face-on optical depth of $\sim 10^{-4}$ primarily show an azimuthally-symmetric ring at 40--47 AU in submillimeter and infrared wavelengths; this ring is associated with the cold classical Kuiper Belt.   For models with lower optical depths ($10^{-6}$ and $10^{-7}$), synthetic infrared images show that the ring widens and a gap opens in the ring at the location of of Neptune; this feature is caused by trapping of dust grains in Neptune's mean motion resonances.
At low optical depths, a secondary ring also appears associated with the hole cleared in the center of the disk by Saturn.  Our simulations, which incorporate 25 different grain sizes, illustrate that grain-grain collisions are important in sculpting today's Kuiper Belt dust, and probably other aspects of the Solar System dust complex;  collisions erase all signs of azimuthal asymmetry from the submillimeter image of the disk at every dust level we considered.  The model images switch from being dominated by resonantly-trapped small grains (``transport dominated") to being dominated by the birth ring (``collision dominated") when the optical depth reaches a critical value of $\tau \sim v/c$, where $v$ is the local Keplerian speed.    
 
\end{abstract}


\section{Introduction}

Debris disks are often described as more massive versions of the Solar System's Kuiper Belt \citep[e.g.][]{grea04, bryd06, jewi09, boot09}.   Debris disks, like the disks around Fomalhaut, Vega, $\epsilon$ Eridani, etc. can only be imaged via existing techniques if they have optical depths of $\sim 10^{-4}$ or higher.   Models of Kuiper Belt (KB) dust production informed by dust detectors in the outer Solar System suggest a face-on optical depth of more like $10^{-7}$ for the KB \citep{back95, ster96, yama98}.   But perhaps when the KB was younger and more massive, it closely resembled the debris disks we have seen so far around other stars.

This analogy has many ramifications.  For example, images of debris disks around nearby stars show rings, clumps, warps and other asymmetries; these asymmetries have often been compared to the asymmetries in the Kuiper Belt caused by dynamical perturbations from Neptune and other planets.   When we see these patterns in debris disks, can we recognize the planets that are sculpting them?   Can we use the patterns to find hidden planets that we couldn't otherwise detect, or measure the orbital parameters of planets orbiting too slowly to track?    The Kuiper Belt, because of its proximity to the Earth, is potentially an important laboratory for testing our dynamical models of debris disks and our ideas about debris disk morphologies.

Several authors have made dynamical models of the distribution of dust in the Kuiper Belt for comparison with images of other debris disks.   \citet{lz99} showed that Neptune may temporarily trap dust in mean motion resonances (MMRs), forming a wide circumsolar ring, from 35--50 AU, with a gap in the ring at the location of Neptune.   This model has often been compared to the wide, clumpy rings seen around Epsilon Eridani and Vega \citep[e.g.][]{maci03}.   \citet{mm1} explored how grains of various sizes behave in the outer Solar System, and predicted the spectral energy distribution of the Kuiper belt dust \citep[see also][]{mm2}.   \citet{holm03} explored how a particular family of KBOs, the plutinos, could contribute to the resonant Kuiper Belt dust population.

But these models contain an important limitation: they largely neglect grain-grain collisions.
For some grain sizes in any debris disk, the typical collision time becomes shorter than the typical Poynting-Robertson (PR) time, affecting the disk morphology \citep[e.g.][]{wyat05}; we find that this effect sets in at even lower optical depths than previously anticipated.   Moreover, as we mentioned above, the debris disks we see around other stars are much more massive than the KB, making collisions even more important; virtually all known extrasolar debris disks show ``collision-dominated'' behavior.   A recent paper by \citet{vite10} illustrates the importance of these grain-grain collisions on the Kuiper belt dust distribution, thought it does not model the dynamical effects of the planets.

   
In this paper, we take a step toward a better understanding of the analogy between the KB and extrasolar debris disks.  We use our new ``collisional groming" algorithm \citep{star09} to explore the effects of grain-grain collisions and planetary perturbations together on the distribution of Kuiper Belt dust.  We break the KB source population into three populations: hot, cold and plutinos.  We model the effects of grain-grain collisions in today's KB, and we model how the KB dust morphology would change if the amount of dust were increased from a face-on optical depth of $\sim 10^{-7}$ to $\sim 10^{-6}$, $\sim 10^{-5}$, and $\sim 10^{-4}$.     

   

\section{Numerical Techniques: Collisional Grooming}
\label{sec:numerical}

Here is a brief summary of the collision grooming algorithm. \citet{star09} described the algorithm in depth and various numerical tests it has passed.   First the orbits of a set of dust grains are numerically integrated using an n-body integrator, and the positions and velocities of the particles are recorded periodically in a histogram.  We call this histogram the ``seed" model.  Then the trajectories of each particle are re-interpreted as steady state streams of particles, with weights that define the number of particles in each stream at any given point along the trajectory.   The weights are then iteratively manipulated so that they describe a self-consistent cloud of interacting particles.  The result is a 3-D grid that contains the number density of the cloud, a self-consistent solution to both the dynamical equations that govern the particle trajectories and the number flux equation that accounts for the creation and destruction of particles in every histogram bin.

There have been several recent papers on kinetic treatments of collisions in debris disks \citep[e.g.][]{kriv06, wyat07} including the Kuiper Belt \citep{kriv05}.   Some of these models involve more detailed collision physics than our simulations, e.g., time evolution and fragmentation.  But our simulations have the unique capability to explore the interaction between the dynamical effects of planets, such as resonances, secular forcing, etc., and grain-grain collisions.   These pheneomena are crucial for understanding the distribution of dust in debris disks, even in the presence of collisions, as we will show.

To create the seed model for our study, we integrated the equations of motion for a particle subject to gravity from the Sun, Jupiter, Saturn, Uranus, and Neptune plus radiation pressure, Poynting-Robertson (PR) drag and solar wind drag \citep[see][]{star08}.    We used a customized hybrid symplectic integrator, described in \citet{star08}, modified to include drag forces.  The integrations all used a symplectic time step of 0.1747 years, equal to $\sim$ 1/20th of the orbital period at 2.5 AU.  We ran the integrations on NASA's Discover cluster.    We recorded the seed model in the frame rotating with the mean motion of Neptune, to capture any resonant structures associated with that frame \citep[see][for a discussion of which resonant structures are associated with which frame]{kuch03}.


We integrated the orbits of all particles for $5 \times 10^8$ years, a few times the maximum collision time for any particles we considered.  We removed particles once they reached a semimajor axis $a < 2.5$ AU, or $a > 300$ AU, or suffered a collision with a planet, assuming realistic planet radii.  We chose the time between records individually for each particle size bin, such that each size bin contributed roughly $4 \times 10^{6}$ records for each of the three source populations (see below).  We accumulated the particle records in a histogram of 512 $\times$ 512 $\times$ 128 bins, each with size 0.5 $\times$ 0.5 $\times$ 0.3 AU.   We ran the collisional grooming algorithm for as many iterations as it took until none of the weighting factors varied by more than 10\% from one iteration to the next.   This part of the simulation took only a few hours on a single processor, but required 20 gigabytes of of RAM.

\citet{mm1} pointed out that with too few particles, it is possible that 1) the mean motion resonances (MMRs) may not be populated accurately, and that 2) a few unusually long-lived grains can dominate the simulations.    We overcome these sources of noise in the same manner as \citet{star09}.  We ensure the MMRs are populated accurately by using a total of 75,000 particles; \citet{mm1} had called for $\sim 10^5$.   We handle long-lived particles using the collisional grooming algorithm, which includes the effects of collisions in removing these particles.

We added a new physical detail to our simulations since \citet{star09}.  In our new models, when two dust grains collide, they only destroy each other when the energy of the collision measured in the center of mass frame exceeds the estimated binding energy of the particles, $Q m_t$, where $m_t$ is the mass of the target grain.   Otherwise, the particles continue unaltered.  We use an estimate for the specific binding energy described in \citet{kriv06}, the ``strength" regime of Equation~22 in that paper:
\begin{equation} 
Q = A_s  \left({s \over {1 \ {\rm m} }} \right)^{b_s}  
\end{equation} 
The radius of the dust grain is $s$.  We take $A_s = 2 \times 10^{5}$ erg g${}^{-1}$ and $b_s = -0.24$, the values \citet{kriv06} used for ``icy'' grains.   Throughout this paper, we assume spherical particles with a density of $\rho= 1$ g cm${}^{-3}$.   

For the sake of numerical simplicity in this first generation of 3-D multi-grain-size collisional models, we do not explicitly follow any fragments produced in the collisions; we assume that all dust production arises in the source populations, described below.    In any case, samples of cometary particles directly returned from the Stardust mission \citep{brow06, zole06} and  observations of cometary ejecta during the Deep Impact mission \citep{ahea05} reveal that the majority of  observed cometary particles are loosely bound aggregates of submicron-sized grains, which can easily be shattered into unbound $\beta$-meteoroids.     These samples seem likely to represent KB particles too.

\section{Source Populations}
\label{sec:initial}

In the models described here, there are two kinds of bodies:  dust grains, and source bodies.   The dust grains have orbits that evolve via drag and radiation pressure; they can be destroyed in collisions with each other.   The source bodies have fixed orbits, and steadily produce the dust grains; they model bodies too large to be destroyed in collisions, but which nonetheless release dust, e.g., via collisions that are not explicitly part of the bookkeeping.  

The dust grains contribute most of the optical depth, so we focus mostly on their dynamics.  The large bodies are incorporated into the simulations as the initial conditions of the grains in the seed model.   The models ultimately depict a steady-state flow of grains; ``initial" here refers only to where the individual grains are launched in the seed model.  

The size of a grain is approximately parametrized by $\beta$, the force on the grain from radiation pressure divided by the force from stellar gravity:
\begin{equation}
\beta = {{3 L_{\star} Q_{\rm PR}} \over {16\pi G M_{\star} c \rho s}}
\label{eq:beta}
\end{equation}
where $Q_{\rm PR}$ is the radiation pressure coefficient.  \citet{lz99} used four different $\beta$ values and a total of 350 particles.  \citet{mm1} used five different $\beta$ values and a total of 500--700 particles, for each of four different models.   We used 25 different $\beta$ values and a total of 75,000 particles.    The 25 $\beta$ values range from 0.00046 to 0.43355; the spacing between them is logarithmic.   Since we assume perfectly absorbing spherical particles with a density of $\rho =1$ g cm${}^{-3}$, $\beta = 0.57 \ \mu \rm{m}/s$, where $s$ is the dust grain radius.  With this assumption, the range of sizes in our initial conditions corresponds to 1.3 to $1239  \ \mu$m.

The grains were launched with a size distribution $dn/ds = s^{-3.5}$, where $s$ is the radius of the grain.  This distribution is the ``crushing law" telling us the relative production  rate of grains of various sizes.   We discuss how collisional processing alters this size distribution below.

Our study benefits from the recent explosion in KBO surveys.  Our simulations incorporated three different populations of source bodies, representing three different populations of KBOs.   We relied on the models of \citet{kave08} and \citet{kave09} to disentangle these populations in the face of the many observational biases that affect measurements of KBO populations \citep[see also][]{brow01, truj01}.   The source populations we assumed are as follows:

\begin{itemize}

\item {\bf Cold.}
This source population represents the cold classical Kuiper Belt \citep{brow01}.  The semimajor axes, $a$, for this population were distributed uniformly between 42.5 and 45 AU.    The eccentricities, $e$, were distributed uniformly between 0 and 0.1.   The inclinations, $i$, were distributed with distribution $P(i) \propto \exp(-0.5 (i / \sigma_C)^2)$ where $\sigma_C=1.5^{\circ}$.
The longitudes of ascending node and arguments of perihelia were distributed uniformly over $[0,2\pi)$.  This component makes up 16.3 \% of the total source population.

\item {\bf Hot.}
This population represents both the hot population of the classical KBOs and the scattered/detatched Kuiper Belt.    The semimajor axes are distributed uniformly between 35 and 50 AU, and the eccentricities are distributed such that $P(e) \propto e$, subject to the additional criterion that $a(1-e) > 35$ AU.  Because of this additional criterion, the semimajor axis distribution ends up weighted toward 50 AU.  The inclinations are distributed with $P(i) \propto \exp(-0.5 (i / \sigma_H)^2)$ where $\sigma_H=13^{\circ}$.  The longitudes of ascending node and arguments of perihelia were distributed uniformly over $[0,2\pi)$.  The dominance of this category of object in the KB has only recently become apparent \citep[e.g.][]{truj00}; we assumed it makes up 79.7 \% of the total source population.

\item {\bf Plutinos.}
To represent these bodies, we chose the orbits for the source bodies from a list of orbits for actual KBOs on the Minor Planet Center's web site with $39.1 < a < 39.7$ AU.  We assumed that this population makes up 4\% of the total source population.

\end{itemize}

\begin{figure}[!ht]
\centerline{
\includegraphics[angle=0,height=4in]{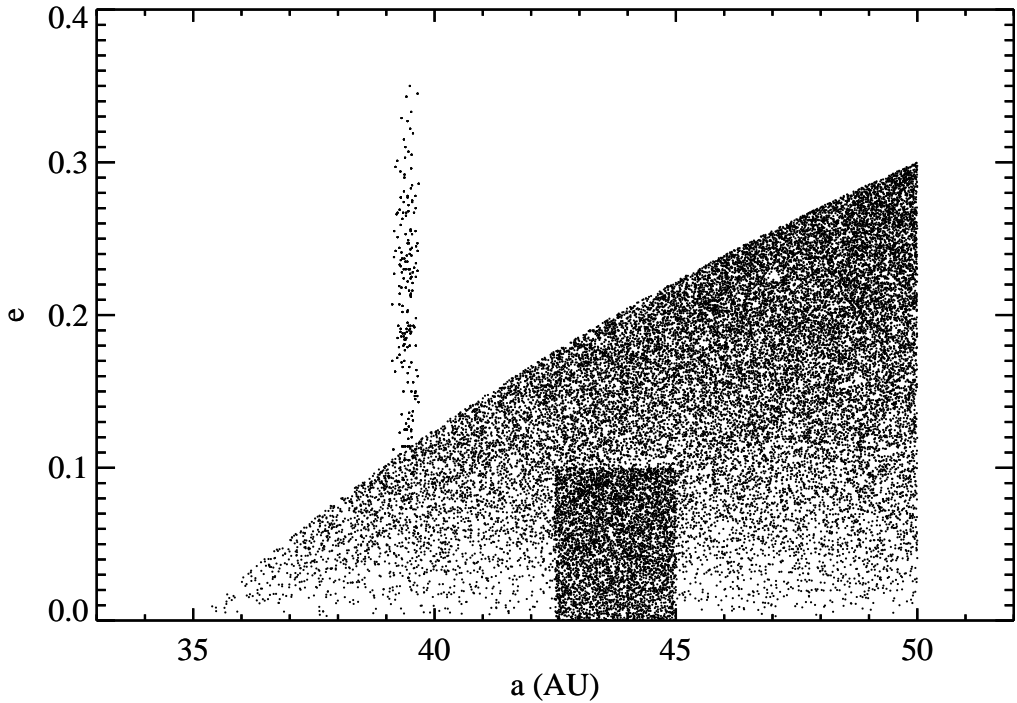}
}
\caption{Semimajor axis/eccentricity distribution for all the assumed source particles.}
\label{fig:initial}
\end{figure}

Figure~\ref{fig:initial} illustrates these three assumed source populations.  We assigned 25,000 particles to each of them.  For comparison, \citet{lz99} assumed that all of their source bodies were in orbits with semimajor axes 45 or 50 AU. \citet{mm1} assumed all source bodies had orbits with semimajor axes equal to 45 AU, or that the source body semimajor axes were distributed uniformly from 35--50 AU.   \citet{holm03} assumed all their source bodies had approximately Pluto-like orbits.  Note that although we assigned equal numbers of particles to each source population, we assigned each dust population a different relative dust production rate in the collisional grooming algorithm, as described above (16.3\% cold, 79.7\% hot, 4\% plutinos).

Many of the source particles have resonant orbits, by chance.   But except for the plutinos, we did not attempt to capture the detailed resonant dynamics of the KB in our source bodies \citep[see][]{chia03}.  When the grains in our models are released, they instantly jump to new orbits because of radiation pressure, conserving their velocity at release \citep[see][]{mm1, holm03}.   Creation of particles through collisions generates some additional velocity dispersion \citep[e.g.][]{cell99}, though we do not attempt to explicitly model this effect.  These two effects will often wash out the resonant behavior of the source particles; see \citet{holm03}.   The KB's detailed resonant structure might serve to enhance the resonant populations of  dust beyond what our models show, but we leave an investigation of this effect for a future date when we understand this phenomenon better observationally.

For our basic KB model, we chose a total dust production rate 
of $3.6 \times 10^6$ g s${}^{-1}$ to make the maximum face-on
geometric optical depth in the ring $\sim 10^{-7}$.   This rate is
consistent with estimates based on the dust fluxes measured by
{\it Pioneer 10} and {\it 11} beyond 10 AU from the Sun \citep{land02, mm2}.
We also ran models where we scaled the dust production rate up to
increase the maximum face-on total optical depth ($\tau_{\rm{max}}$) at 42 AU to $\sim 10^{-6}$, $\sim 10^{-5}$, and $\sim 10^{-4}$.    


Table~\ref{tab:criticalgrainsize} lists the geometric face-on optical depths at 42 AU and dust production rates, summed over all grain sizes in the model, for each of the four dust levels we considered.   The dust production rate depends on the size range of grains considered---especially the size of the largest grains considered.   So the table quotes both the total dust production rate and the rate of production of grains with $s < 10 \ \mu$m, for ease of comparison with other calculations.  For example, \citet{yama98} estimated that the total dust production rate for particles smaller than $10 \ \mu$m  is (0.37Ð-2.4) $\times 10^6$ g s${}^{-1}$ if the objects have hard icy surfaces, or (0.85Ð-3.1) $\times 10^7$ g s${}^{-1}$ if the objects are covered with icy particles smaller than interstellar grains.   The hard icy surfaces case lies between our  $\tau_{\rm{max}} \sim 10^{-7}$   model and our $\tau_{\rm{max}} \sim 10^{-6}$ model, while the small icy particles case lies between our $\tau_{\rm{max}} \sim 10^{-6}$ model and our $\tau_{\rm{max}} \sim 10^{-5}$ model.  The \citet{vite10} model indicates an even higher dust production rate, corresponding to $\tau_{\rm max} \approx 2 \times 10^{-5}$.

\begin{deluxetable}{lccc}
\tablewidth{5.5in}
\tablecaption{Simulation Parameters}
\startdata
\tableline
\tableline
Optical  &  Total Dust  Pro-       &  Dust $<10 \ \mu$m  Pro-                                & Critical Grain  \\
Depth ($\tau_{\rm max}$) &   duction Rate  (g s${}^{-1}$)   & duction Rate  (g s${}^{-1}$)     & Size, $s_{c}$, ($\mu$m$)\tablenotemark{a}$  \\
\tableline
 $\sim 10^{-7}$   &   $3.6 \times 10^{6}$  &   $2.2 \times 10^{5}$   &   17  \\
 $\sim 10^{-6}$    &  $7.5 \times 10^{7}$   &   $4.6 \times 10^{6}$  &  6  \\
 $\sim 10^{-5}$     & $2.4 \times 10^{9}$  &   $1.5 \times 10^{8}$   &  3  \\
 $\sim 10^{-4}$     & $1.2 \times 10^{11}$  &   $7.3 \times 10^{9}$   &  $\sim 1$  \\
\tableline
\tableline
\enddata
\tablenotetext{a}{Grain size where $t_{\rm coll} =  t_{\rm PR}$, as measured in simulation.}
\label{tab:criticalgrainsize}
\end{deluxetable}


\section{Results}
\label{sec:results}

\subsection{Collisionless Simulations}
\label{sec:collisionless}

\begin{figure}[!ht]
\centerline{
\includegraphics[angle=0,height=6in]{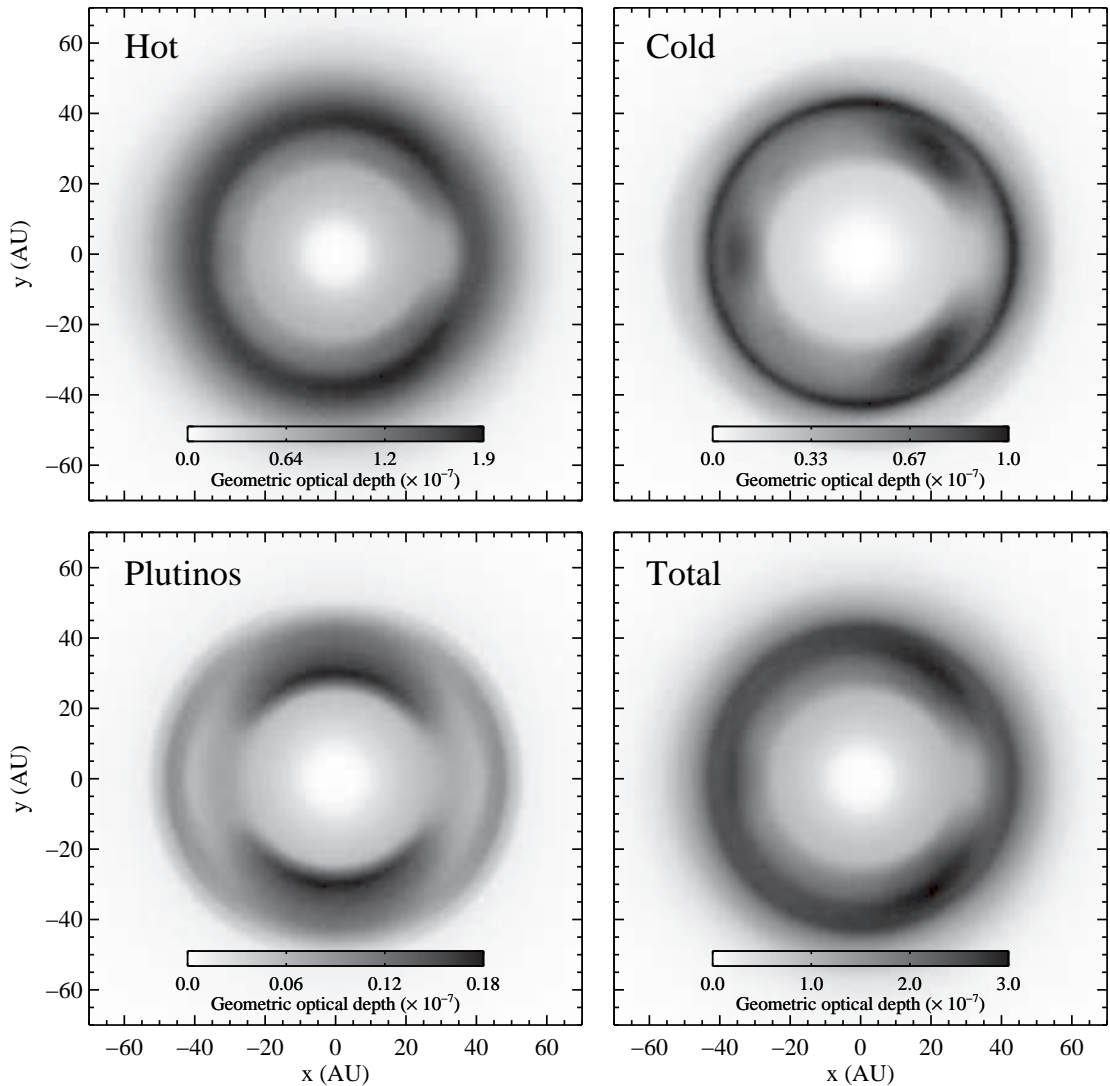}
}
\caption{Collisionless simulations of the Kuiper Belt dust: the geometric optical depth for each source population, and for the total.   Neptune is located at x=30.0696, y=0 AU.}
\label{fig:collisionless}
\end{figure}

Figure~\ref{fig:collisionless} illustrates the seed model we used: a histogram representing the steady-state distribution of KB dust grains in the absence of collisions.  It also shows the contributions to this seed model from each of the three source populations described above.   The figure shows only a 2-D projection of the cloud density; the full seed model is a 3-D histogram, which also contains the 3-D velocity distribution at each point in the histogram.   The grains of different sizes were combined together weighted to simulate a dust production rate of $dN/ds \propto s^{-3.5}$.   The total dust production rate of the model shown in Figure~\ref{fig:collisionless} is the same as that in the $\tau_{\rm{max}} \sim 10^{-7}$ collisional simulation.

This figure can be compared to other collisionless models of the KB dust, like those described in \citet{lz99} and \citet{mm1}.   Overall, the seed model morphology is a wide circumsolar ring with a gap at the location of Neptune, not dissimilar from that predicted by those authors.  It also resembles the Type I resonant ring described by \citet{kuch03}, the case of the low-mass planet on a circular orbit.

In this collisionless stage of the simulation, particles of all sizes participate strongly in resonant trapping, especially in the lowest-order resonances.  The signature of the plutino dust is a ring of dust trapped in the 3:2 MMR with Neptune---like the plutinos themselves.     Dust produced by the cold population of source bodies shows heavy signatures of several MMRs with Neptune:  3:2, 7:4, 8:5, etc., but not the 2:1, because it is released substantially interior to the 2:1.  Dust produced by the hot population also participates in the MMRs, though less than the other populations because of the reduced trapping probabilities associated with higher $e$ and $i$.  

\subsection{Simulations with Collisions}
\label{sec:withcollisions}

\begin{figure}[!ht]
\epsscale{0.8}
\centerline{
\includegraphics[height=6in]{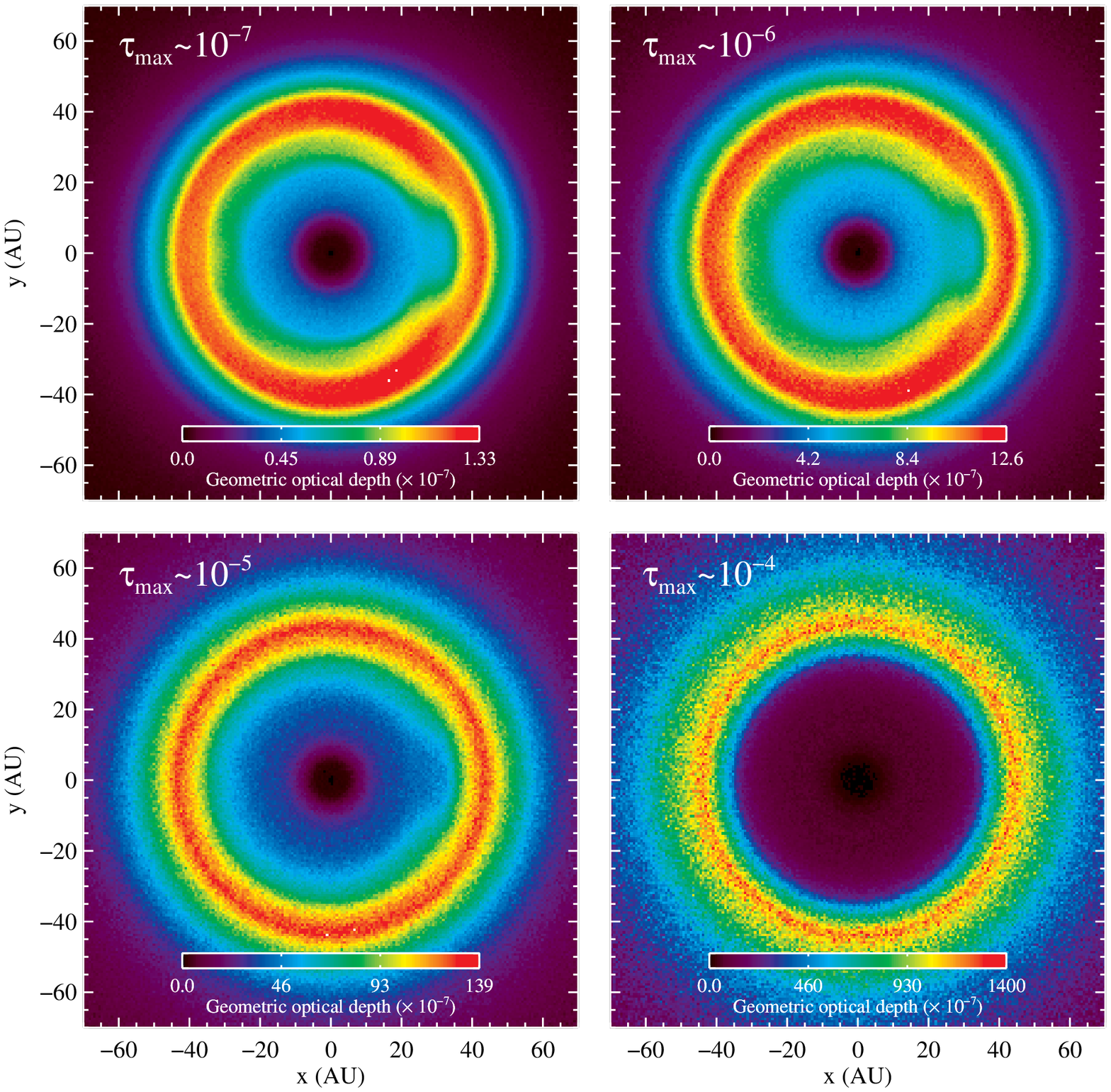}
}
\caption{Simulations with grain-grain collisions: the total geometric optical depth including all sources.  Neptune is located at x=30.0696, y=0 AU.  Each panel represents a different dust level.}
\label{fig:collisions}
\end{figure}

Figure~\ref{fig:collisions} shows the geometric optical depth of the total KB dust population after the full collisional grooming algorithm has been applied, as described above, at four different dust levels.   As the optical depth increases, collisions remove grains from the center of the disk.     Moreover, as \citet{star09} showed, the highest grain-grain collision rates occur in MMRs, so in the absence of resonant parent bodies, the collisions also tend to reshape and then erase the resonant structures.

At the highest optical depth, the pattern mostly resembles a narrow ring, coincident with the cold classical Kuiper Belt.

\begin{figure}[!ht]
\epsscale{0.8}
\centerline{
\includegraphics[angle=0,height=6in]{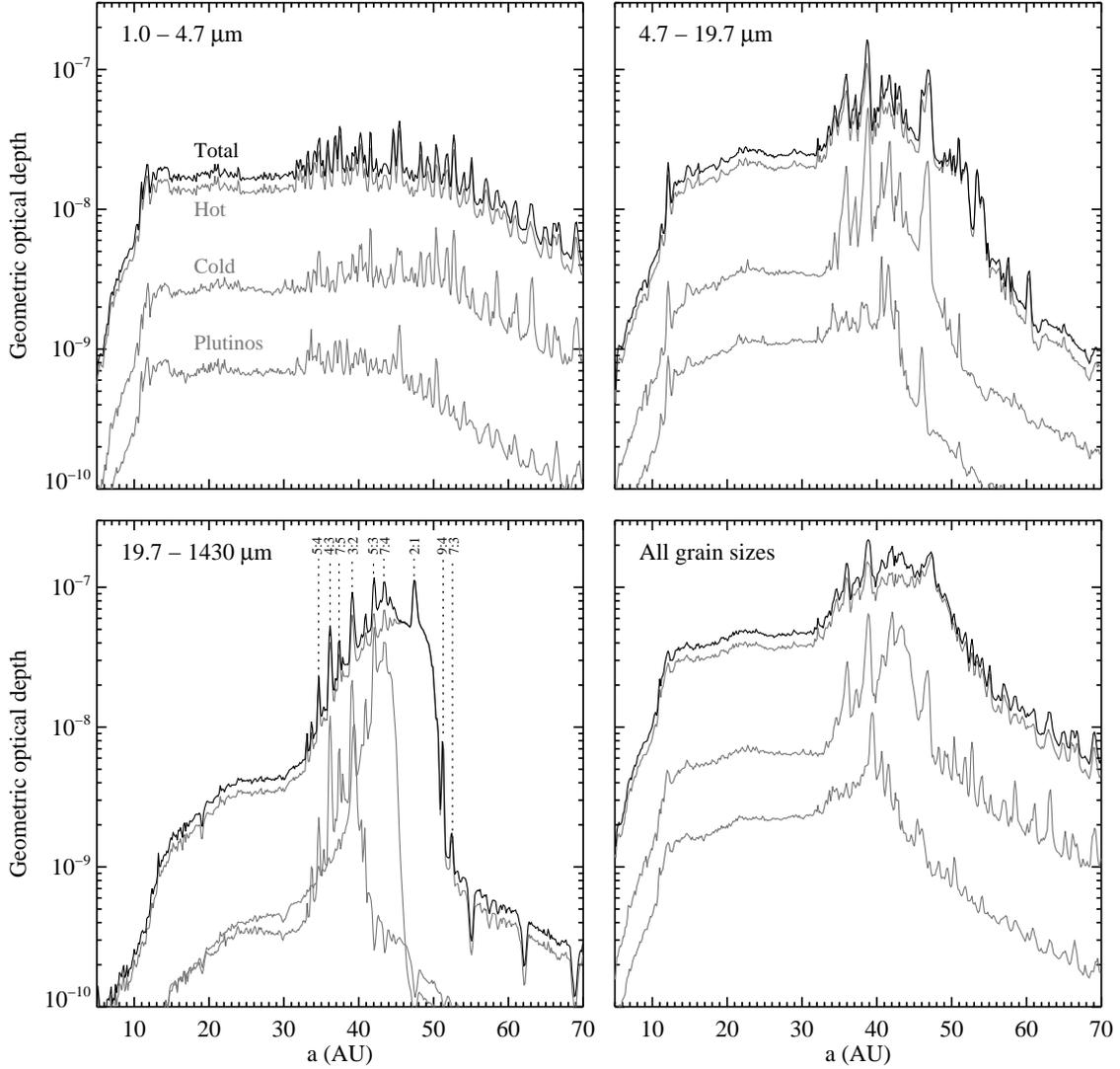}
}
\caption{Semimajor axis distributions of particles in the $\tau_{\rm{max}} \sim 10^{-7}$ collisional simulations, according to grain size.  Grey curves show the contributions from the individual source populations.  Dashed lines label some mean motion resonances with Neptune in the bottom left panel. }
\label{fig:semimajoraxis1}
\end{figure}

\begin{figure}[!ht]
\epsscale{0.8}
\centerline{
\includegraphics[angle=0,height=6in]{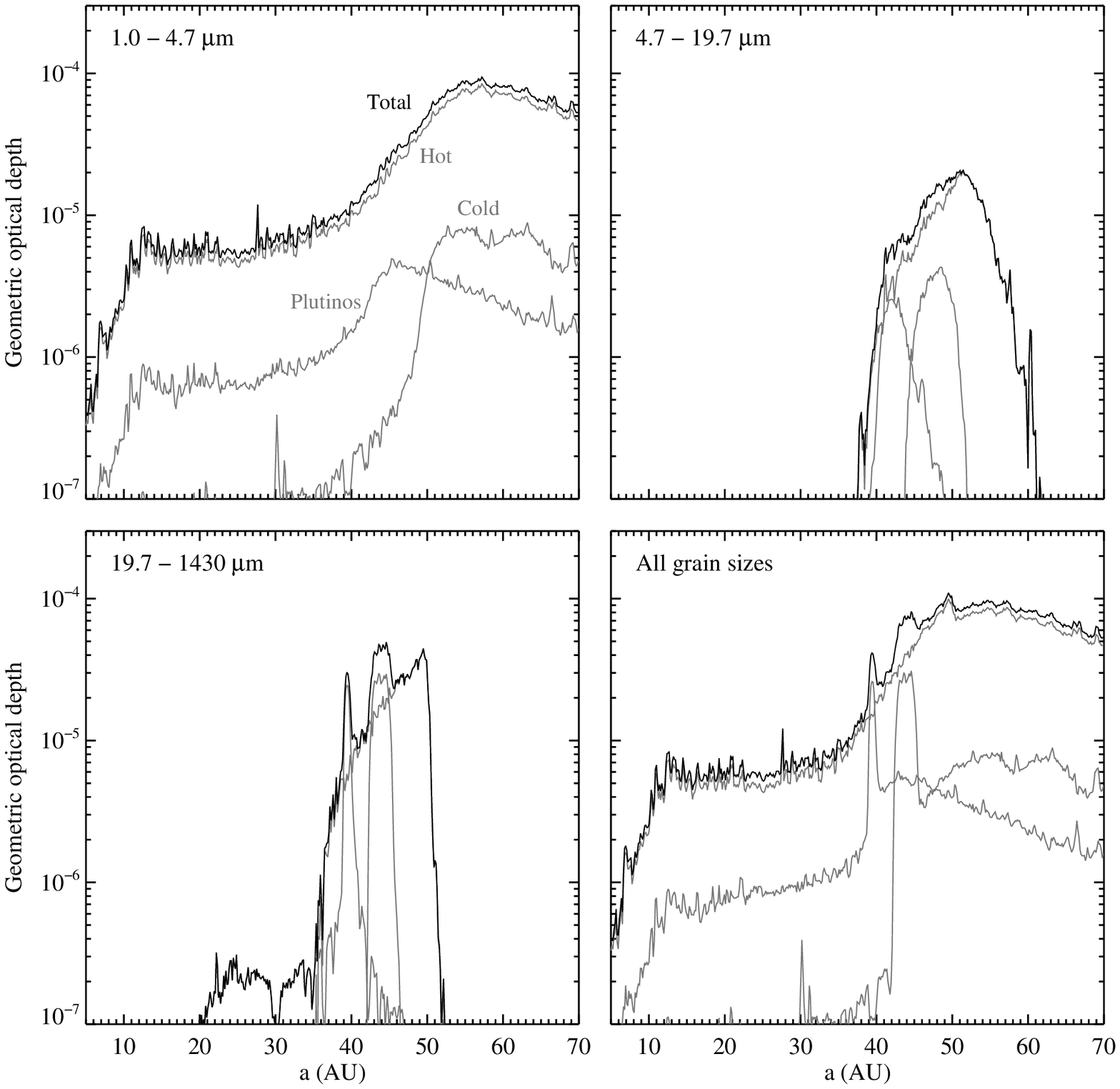}
}  
\caption{Semimajor axis distributions of particles in the $\tau_{\rm{max}} \sim 10^{-4}$ simulations. }
\label{fig:semimajoraxis1000}
\end{figure}

Figures~\ref{fig:semimajoraxis1}, and \ref{fig:semimajoraxis1000}  show the semimajor axis distributions of the grains in the four collisional simulations, combined into three size bins (it would be impractical to show all 25 size bins).   It also shows the distributions summed over all grain sizes. To calculate the ``optical depth" shown in in these figures, we took the number of grains within each semimajor axis bin, multiplied by the grain cross section, and divided by $2 \pi a \Delta a$.   The bins have width $\Delta a = 0.14$ AU   The grey lines show how the distributions break down by source population.  The black lines show the total in each size bin.

One phenomenon that these figures show immediately is that only small grains penetrate interior to Neptune's orbit.  The result is a disk with large grains on the outside, small on the inside, like the disk around Fomalhaut, according to data from {\it Spitzer} \citep{stap04} and VLTI/VINCI  \citep{absi09}.  This phenomenon occurs even for a disk with maximum optical depth $\sim 10^{-7}$.

Let us look more closely at which grain sizes dominate the optical depth throughout the disk.  At low collision rates ($\tau_{\rm{max}} \sim 10^{-7}$), the 4.7--$19.7 \ \mu$m particles dominate the semimajor axis distribution interior to about 50 AU.  At $\tau_{\rm{max}} \sim 10^{-6}$ (not shown), that bin and the bin with the smallest grains (1--$4.7 \ \mu$m) present roughly equal contributions at $a < 50$ AU.   At $\tau_{\rm{max}} \sim 10^{-4}$, the smallest grains dominate everywhere but 35--50 AU, where the largest grains make a comparable contribution (Figure~\ref{fig:semimajoraxis1000}).    We will discuss this shift in the dominant grain size, from larger to smaller as the optical depth increases, further below.

Now let us look at the many fine peaks in the semimajor axis distributions of the dust grains; these peaks arise mainly from dust trapped in MMRs.  To read these peaks, it is helpful to remember that the mean semimajor axis in a MMR is shifted inward for dust grains by a factor of $(1-\beta)^{1/3}$ because radiation pressure counteracts stellar gravity, decreasing the orbital period for small grains.   For the $5 \ \mu$m grains in our simulations, the inward shift amounts to about 1 AU in the KB.  Over the range of 1--$5 \ \mu$m, the shift ranges over a factor of several AU; this probably accounts for the many fine peaks in the semimajor axis distribution of the 1--$4.7 \ \mu$m grains.   The duplicate and split peaks in the 1--$4.7 \ \mu$m grain semimajor axis distributions are an artifact of the quantized grain size distribution used in the simulations.

Nonetheless,  notice how the textures of almost all the curves change at about 30 AU.  Interior to $\sim 30$ AU, the curves in Figure~\ref{fig:semimajoraxis1} are relatively smooth.   Exterior to this semimajor axis, the curves become ragged and bumpy.   This texture reveals the presence and population of many Neptune MMRs, both low order (like 3:2 and 4:5) and higher order (like 8:5, 7:5, and 7:4).  The opposite effect occurs in Figure~\ref{fig:semimajoraxis1000}.   In this Figure, the ragged texture, indicating MMRs, exists only interior to about 30 AU.

In the low-optical depth simulations, several MMRs stand out as having particularly strong peaks.  On the left side of Figure~\ref{fig:semimajoraxis1}, the small grains show a strong peak near the 3:2 MMRs with Saturn ($\sim 12$ AU) and a secondary peak near the 2:1 ($16$ AU).   Very few grains of any size make it interior to 10 AU; they are scattered out of the Solar System by Saturn.

A series of exterior MMRs with Neptune shows strong peaks at all grain sizes in the $\tau_{\rm{max}} \sim 10^{-7}$ model: 4:3 3:2 7:4 2:1 ($\approx$36, 39, 44 and 48 AU).   Dashed lines show these and a few other MMRs in Figure~\ref{fig:semimajoraxis1}.  Some of these peaks also survive in the total semimajor axis distribution.  They appear in all three source populations, though they are strongest in the cold classical population, probably these objects have small eccentricities and inclinations, making them easier to trap in MMRs.  Though the plutino dust is released from bodies in Neptune's 3:2 MMR, only grains larger than $19.7 \ \mu$m remain tightly concentrated around that MMR, at $\approx 39$ AU.    Figure~\ref{fig:collisions} shows that at this dust level ($\tau_{\rm{max}} \sim 10^{-7}$), the resonant structure looks like a ring at Neptune's orbit with a gap, not too dissimilar from that found by \citep{lz99} and \citep{mm1}.  
   
As the optical depth increases, the resonant patterns change.   Figure~\ref{fig:collisions}) shows that the geometric optical depth in the $\tau_{\rm{max}} \sim 10^{-6}$ simulation looks again like a ring with a gap, but narrower.  This pattern indicates that there still are MMRs near Neptune populated with dust. 
In the $\tau_{\rm{max}} \sim 10^{-5}$ simulation, however, the ring becomes more azimuthally-symmetric, and a transition occurs, which we will discuss below.

Figure~\ref{fig:semimajoraxis1000} shows a very different disk than Figure~\ref{fig:semimajoraxis1}.    Here, the peaks  in the MMRs vanish beyond $\sim 30$ AU.  The only strong resonant signatures that persist in the Kuiper Belt region are in the large ($> 19.7 \ \mu$m) grains from the plutinos.  Some small grains in the interior of the disk also appear to be trapped in MMRs with Saturn.

\subsection{Grain Size Distributions}
\label{sec:grainsize}

\begin{figure}[!ht]
\epsscale{0.8}
\centerline{
\includegraphics[angle=0,height=3.5in]{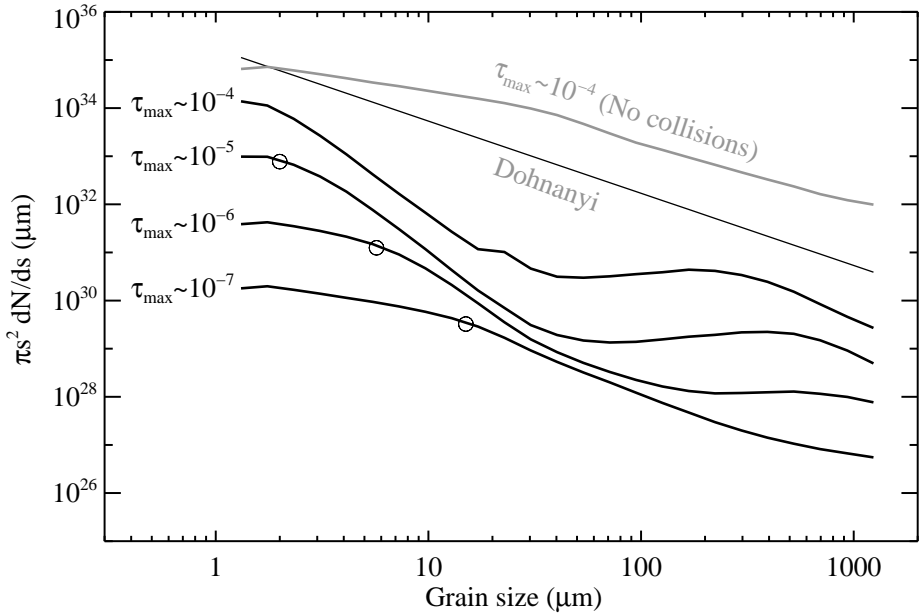}
}
\caption{Number of grains per size interval times times $\pi s^2$.  Black lines show collisional simulations with optical depths of $10^{-7}$, $10^{-6}$,  $10^{-5}$, and $10^{-4}$.  Grey lines show the collisionless simulation scaled to $\tau_{\rm max} \sim 10^{-4}$ and a $s^{-3.5}$ power law \citep{dohn69} times $\pi s^2$ for comparison.    Circles indicate the critical grain size, $s_c$ for each simulation, as listed in Table~1.
}
\label{fig:grainsizes}
\end{figure}

Figure~\ref{fig:grainsizes} illustrates the steady-state size distributions of particles, $dN/ds$, in our Kuiper Belt simulations, integrated over the whole cloud.  Specifically, the figure shows $dN/ds$ multiplied by the grain cross section, $\pi s^2$, to show exactly which grain sizes dominate the geometric optical depth.  Black curves show the size distributions for simulations with $\tau_{\rm{max}} \sim 10^{-7}$, $10^{-6}$, $10^{-5}$, and $10^{-4}$.   A grey line shows a power law of $s^{-1.5}$, representing both the crushing law we used, and the \citet{dohn69} collisional equilibrium power law (multiplied by $\pi s^2$).   Another grey line shows a simulation with no collisions, scaled to yield an optical depth in the Kuiper Belt similar to that of the $\tau_{\rm{max}} \sim 10^{-4}$ collisional simulation. 


When we presented the collisionless simulations, above, we weighted the contributions from grains of different sizes so the final size distribution matched a prescribed power law.  The collisional grooming algorithm does not require any such tuning!  In the collisional grooming algorithm, the grains are released according to a crushing law, and the algorithm calculates, self-consistently, the steady-state size distribution of the particles everywhere in the disk.  

Figure~\ref{fig:grainsizes} shows that  in the absence of collisional processing, the slope of the steady-state size distribution becomes shallower than the crushing law we assumed.   This shallower slope represents two effects, described in \citet{star08}.  First of all, the PR time is proportional to $s$, so in the absence of planets or collisions, the size distribution would become one factor of $s$ shallower.  Second, larger grains are more likely to be trapped in MMRs with planets, which prolong their lifetimes.   

When the collisions are turned on, however, the size distribution relaxes to something closer to the \citet{dohn69} distribution.   The average slope of the size distributions for the collisional models is $-3.5$.     The curves for the collisional models also contain dips in the mid-sized grain populations.   The dips become deeper as the optical depth increases.

\subsection{A Critical Grain Size and a Crossover Optical Depth}

\citet{star08} hypothesized that in the absence of a resonant source population, grains with size such their PR time, $t_{\rm PR}$, matched their collision time, $t_{\rm coll}$, would dominate the optical depth of a resonant ring.  Our new simulations seem to confirm this hypothesis.  We will call this critical grain size $s_c$.   Table~\ref{tab:criticalgrainsize} shows $s_{c}$ in each simulation, averaged over the first 50 records in each stream.  The critical grain size is also plotted in Figure~\ref{fig:grainsizes} for each simulation as a circle on the corresponding black curve.

We can estimate $s_c$ if we approximate the collision time as \citep{wyat03}
\begin{equation}
\langle t_{\rm coll} \rangle \sim T_{orbit}/(4 \pi \tau), 
\end{equation}
where $\tau$ is the face-on optical depth.   The PR time is \citep{wyat50}
\begin{equation}
t_{\rm PR} \approx {{4 \pi c^2 s \rho a^2} \over { ({3 L_{\star} Q_{\rm PR}})}},
\end{equation}
where the particle has semimajor axis, $a$, radius, $s$, density, $\rho$, and radiation pressure coefficient, $Q_{PR}$.  Setting $t_{\rm PR} = \langle t_{\rm coll} \rangle$, we find that
\begin{eqnarray}
s_c & \approx& {{ 3 L_{\star} Q_{PR} }\over{ 8 \pi  c^2 \rho \tau G^{1/2}M_{\star}^{1/2}a^{1/2}}} \\
&=& 1145 \ \mu \rm{m}\ Q_{PR} 
\left({{\rho} \over {1 \  \rm{g}\ \rm{cm}^{-1}}}\right)^{-1} 
\left({L_{\star} \over { L_{\bigodot}}}\right) 
\left({M_{\star} \over {M_{\bigodot}}}\right)^{-1/2} 
\left({a \over {1 \ AU}}\right)^{-1/2}
\left({\tau\over {10^{-7}}}\right)^{-1}.
\end{eqnarray}
In our simulations of the KB, this expression overestimates $s_c$ by a factor of 1--10, as you can tell from Table~1, mostly because the collision time is shorter for grains in MMRs.  The degree of overestimation is highest at low optical depths, where resonant trapping is the strongest.

Overall, the semimajor axis distributions of the grains reveal three kinds of behavior among the various grain sizes.  The smallest grains participate relatively little in the resonant trapping; their radial distribution tends to resemble the solutions to the one-dimensional mass flux equation in \citet{wyat05}.   The largest grains have such large PR times that they tend to be destroyed by collisions before they evolve far from the source particle orbits where they are released.    Grains of intermediate sizes, where the PR time, $t_{\rm PR}$, is comparable to the collision time, $t_{\rm coll}$, dominate the resonant peaks in Figure~\ref{fig:semimajoraxis1}. 

Figure~\ref{fig:grainsizes} also shows the importance of $s_c$.   Each of the curves for the collisional simulations in this figure shows a break somewhere between 1 and $20 \ \mu \rm{m}$.     As you can see, each break coincides with one of the circles in the plot that mark $s_c$.      (The break for the $\tau_{\rm{max}} \sim 10^{-4}$ model probably occurs near $1 \ \mu \rm{m}$, where the simulations do not have particles.)  The breaks show where collisions start limiting grain lifetimes to roughly $t_{\rm coll}$. 

In our collisional grooming calculations we tend to find two different extremes of disk structure: disks dominated by rings of dust transported by PR drag and temporarily trapped in resonances (e.g. the $\tau_{\rm{max}} \sim 10^{-7}$ simulation), and disks whose appearance is dominated by the distribution of source particles (e.g. the $\tau_{\rm{max}} \sim 10^{-4}$ simulation).  Sometimes these regimes are respectively called ``transport dominated" and ``collision dominated".  In simulations using a single particle size \citep{star09}, we found that the transition between these regimes occurred when the collision time became roughly equal to the PR time, i.e. $s \sim s_c$.  In these new simulations where the particles span a wide range of sizes, we find that the PR-drag dominated regime persists at some level as long as there are {\it any} particles in the size spectrum with $s \sim s_c$.   

So our new simulations prompt us to write down a new criterion for the boundary between these regimes: a crossover optical depth.   For $\tau \lesssim \tau_r$, PR-drag and resonant trapping of small grains dominates the geometric optical depth of the disk; for $\tau \gtrsim \tau_r$, the source population dominates.   The source population may be resonant in either case.   The crossover optical depth, $\tau_r$, is set by the criterion that all particles must survive both collisions and radiation pressure blowout.  

To find $\tau_r$, we can set $\beta=1/2$ in Equation~\ref{eq:beta} to find the blowout size, $s_{\rm blowout}$:
\begin{equation}
s_{\rm blowout} = {{3 L_{\star} Q_{PR} }\over{ 8\pi G M_{\star} c \rho}}.
\end{equation}
Then, we set $s_c = s_{\rm blowout}$, when $\tau = \tau_r$.  We find that $\tau_r$ is simply the orbital velocity divided by the speed of light.   
\begin{equation}
\tau_r \approx v/c.
\end{equation}
For example, in the Kuiper belt $\tau_r \sim 10^{-5}$.   So for the simulations described in this paper, $\tau \gtrsim \tau_r$ in the KB region for the $\tau_{\rm{max}} \sim 10^{-5}$ and $\tau_{\rm{max}} \sim 10^{-4}$ simulations, and the birth ring tends to dominate in those simulations.  For the $\tau_{\rm{max}} \sim 10^{-7}$ and $\tau_{\rm{max}} \sim 10^{-6}$ simulations, small grains trapped in MMRs tend to dominate the images.    

This comfortingly simple expression for  $\tau_r$ shows that the crossover optical depth decreases with distance from the star.  So in a given disk, the KB region is likely to be collision dominated, while the center of the disk is more likely to be transport dominated.  For example, in the habitable zone of a solar mass star, $\tau_r \sim 10^{-4}$.   In other words, exozodiacal clouds are more likely to show asymmetries from resonant trapping of small grains than colder debris disks that are KB analogs.

The existence of this crossover optical depth has another important consequence for the detection of hot dust disks.   Far interior to a source of grains, the disk must be in the transport-dominated regime; any grain present in that region has been transported there.  Therefore, we know that in the center of a debris disk, the face-on optical depth of a disk can never be greater than $\tau_r$ unless there is a second, central source of grains.    

We observe this phenomenon in our simulations.  Figure~\ref{fig:semimajoraxis1000} shows that the dust near Saturn, far interior to the dust source, reaches a maximum optical depth at a factor of a few less than $\tau=v/c$, even as we turn up the dust production rate.  The critical optical depth near Saturn is $\tau_r \approx v/c \approx 3\times 10^{-5}$.    

In general, if we see an exozodiacal cloud with optical depth $\gtrsim 10^{-4}$, like, e.g., the dust in the central 20 AU of the $\epsilon$ Eridani system \citep{mora04, back09}, we can infer the presence of a second source of dust.   This prediction is consistent with observations for $25 \ \mu$m flux in excess of the stellar photosphere with {\it Spitzer} \citep{hine06, bryd06, lawl09}, which find stars with hot dust with luminositites $L_{IR}/L_{\star} > 10^{-4}$ to be much rarer than stars with cold dust.   \citet{wyat06} offers a list of known stars with evidence for hot dust at $< 10$ AU; we can infer all the stars on this list must have central sources of grains, perhaps asteroid-belt analogs, interior to 10 AU.

\subsection{Observable Phenomena}
\label{sec:observable}

We synthesized images from our multi-grain size collisional models to illustrate what they would look like to various telescopes.   To create these images, we illuminated the grains with solar flux appropriate to their distance from the star, and calculated the scattered light and thermal emission.  We assumed simple generic emissivity laws to account for the poor ability of grains to radiate and absorb photons with wavelengths larger than the grain size:  emissivity $\epsilon = 1$  for wavelengths $\lambda \le 2 \pi s$ and $\epsilon = (2 \pi s/\lambda)^2 $ for $\lambda > 2 \pi s$ \citep{back93}.  We used the tools in ZODIPIC \citep{mora04} to self-consistently calculate the temperatures of the grains, given solar radiation and the emissivities above.   For the $\tau_{\rm{max}} \sim 10^{-4}$ images, we rebinned the simulations by a factor of two in each direction, to average down the Poission noise in the histograms, which increases for short collision times, as collisions remove the small grains.  We blocked the region interior to 5 AU with a software mask.  We did not model the point spread function of any telescope.

\begin{figure}[!ht]
\epsscale{0.8}
\centerline{
\includegraphics[angle=0,height=3.7in]{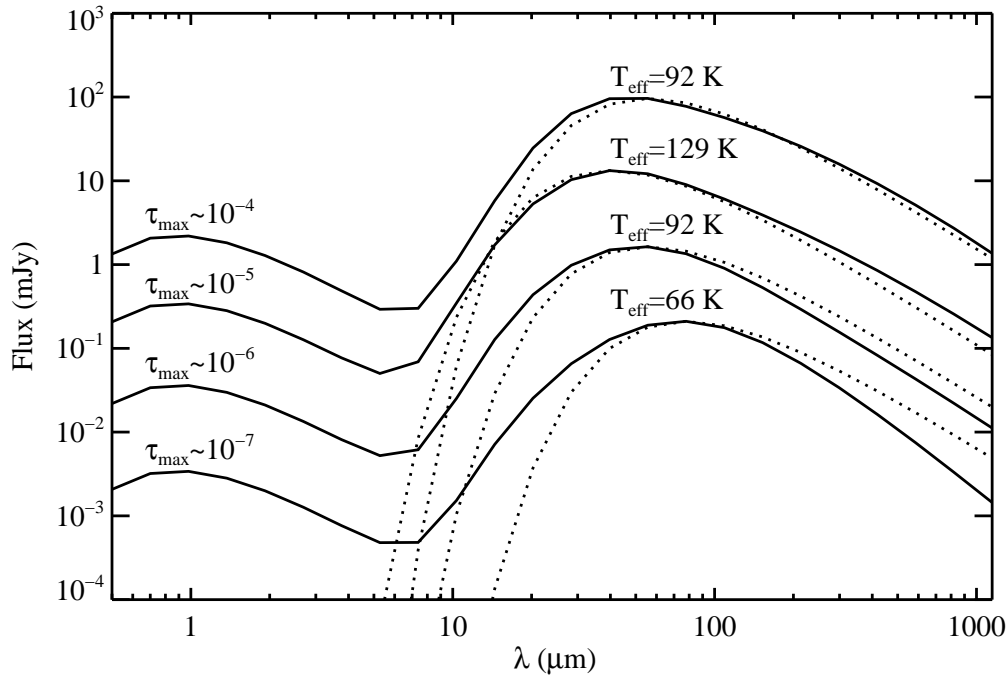}
}
\caption{Spectral Energy Distributions (SEDs) of the four Kuiper Belt dust models at a distance of 10 parsecs, including scattered starlight but not including stellar flux (solid curves).  Dotted lines show blackbody curves with the same peak wavelengths as the models, labeled with their characteristic temperatures.}
\label{fig:seds}
\end{figure}

Figure~\ref{fig:seds} shows the SEDs of the three collisional models, including the contribution from the stellar photosphere.   This figure can be compared to Figure 14 in \citet{mm1}, which does not include the effect of collisions in reshaping the disk.   Dotted lines in the figure show a blackbody curve with the same peak wavelengths as each model, labeled according to their characteristic temperatures.  As the optical depth increases, we find that the peak in the SED from the dust thermal emission moves from $80 \ \mu$m to $40 \ \mu$m as  resonant trapping becomes less effective at retaining grains in the outer Solar System.   As the disk crosses the threshold into the collision-dominated regime, the model curves narrow to more closely resemble the shape of a single-temperature blackbody curve: the emission from large grains in the cold classical Kuiper Belt.  

\begin{figure}[!ht]
\epsscale{0.8}
\centerline{
\includegraphics[angle=0,height=7in]{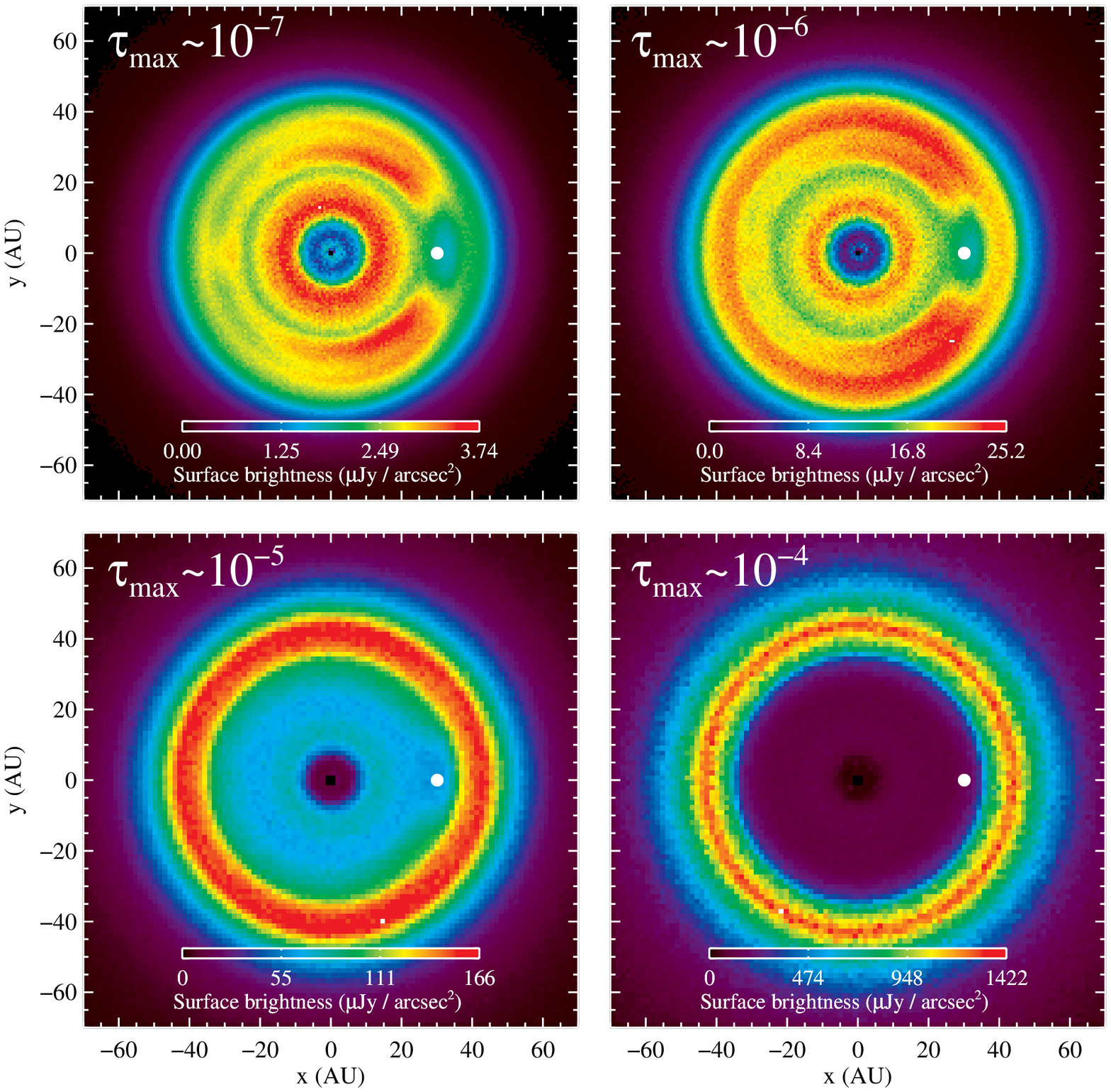}
}
\caption{Images of the collisional dust models at a wavelength of $60 \ \mu$m, near the peak of the thermal emission.  The dot indicates the location of Neptune.  The four models show the appearance of the Solar System assuming Kuiper Belt dust clouds with optical depths of roughly $10^{-7}$, $10^{-6}$, $10^{-5}$, and $10^{-4}$.  A transition occurs where $\tau \approx v/c \approx 10^{-5}$ in the Kuiper Belt.}
\label{fig:ir}
\end{figure}

\begin{figure}[!ht]
\epsscale{0.8}
\centerline{
\includegraphics[angle=0,height=3.7in]{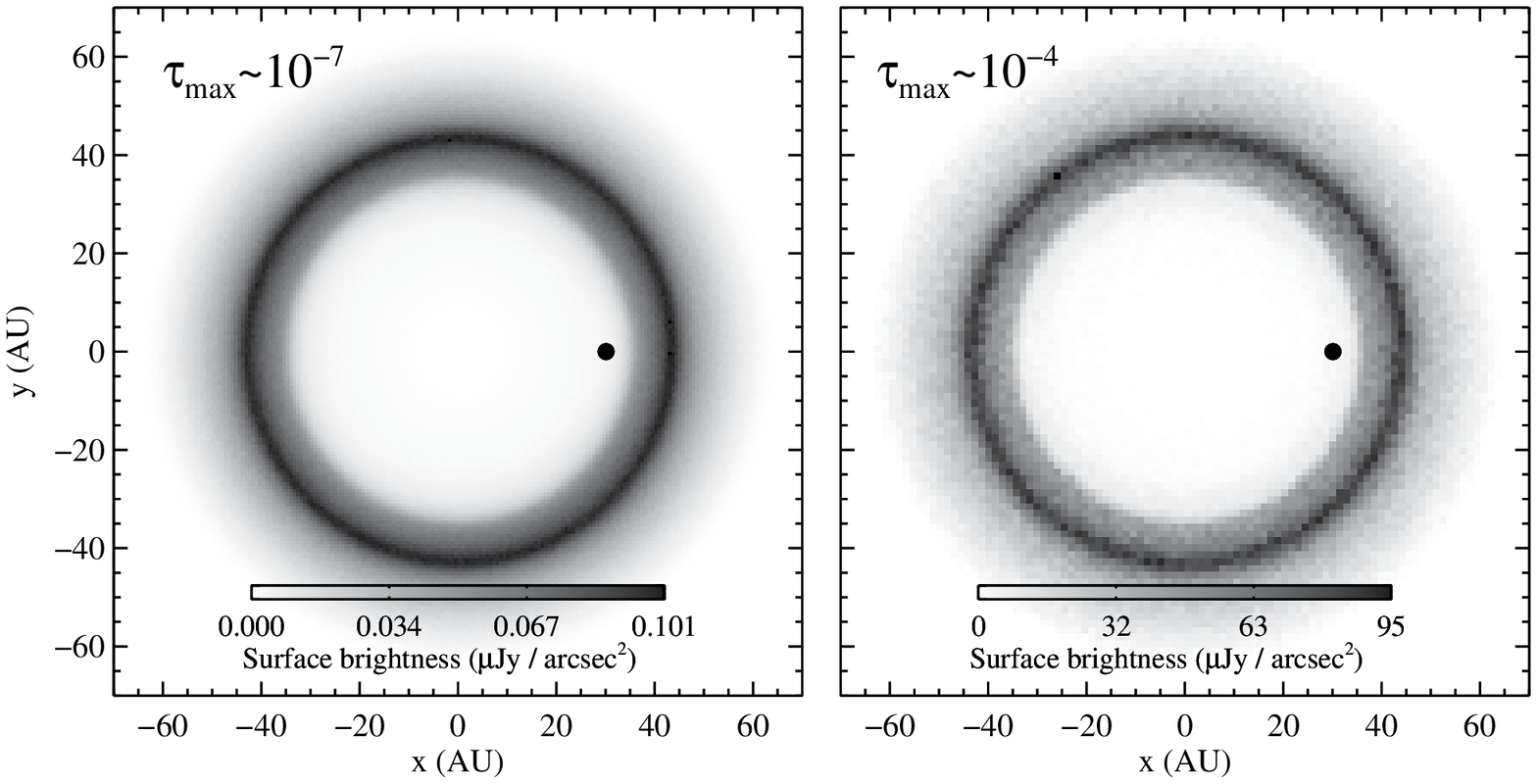}
}
\caption{Images of the collisional KB dust models with optical depths of $10^{-7}$ and $10^{-4}$
at a wavelength of $800 \ \mu$m. The dot indicates the location of Neptune.}
\label{fig:submillimeter}
\end{figure}

Figure~\ref{fig:ir} shows images of the four collisional models viewed at $60 \ \mu \rm{m}$, for comparison with, e.g., images from the Herschel space telescope.   Figure~\ref{fig:submillimeter} shows the $\tau_{\rm{max}} \sim 10^{-7}$ and $\tau_{\rm{max}} \sim 10^{-4}$ simulations as they would appear in the submillimeter ($800 \  \mu$m).   The infrared images show the cloud near the wavelength of its peak emission.  They  emphasize particles with sizes near $60 \ \mu \rm{m}$.   The submillimeter images emphasize larger grains, close in size to $800 \  \mu$m.

Figure~\ref{fig:ir} vividly illustrates the kind of morphological transformation that occurs when $\tau \sim \tau_r$, roughly at an optical depth of $10^{-5}$ in the KB.   At higher optical depths (bottom panels), the images show a narrow ring at 40-47 AU.   At lower optical depths (top panels), a variety of azimuthal asymmetries appear, associated with a ring of dust trapped in MMRs with Neptune.  This ring corresponds to the ring discovered by \citet{lz99}; it has a gap at Neptune's location, indicating the current position of the planet.     We know from Figures~\ref{fig:semimajoraxis1} and~\ref{fig:semimajoraxis1000} that this dust consists of two size ranges:  large grains originating from plutino parent bodies in Neptune's 3:2 MMR, and and smaller grains with $s \sim s_c$, released outside resonance, and then trapped in MMRs.     

The infrared images also show what appears to be a wide ring at 10--20 AU.    We can see from Figure~\ref{fig:semimajoraxis1000} that this ring consists mostly of grains with $s \le s_c$ located just outside Saturn, at an optical depth of $\sim 10^{-5}$.   Some grains are trapped in the 2:1, 3:2, and other MMRs with Saturn.  However, mostly, the feature appears ring-like because grains interior to 10 AU have been scattered out of the Solar System, and the central illumination makes the brightness fall off rapidly with circumsolar distance; the feature is really more like a central hole than a ring.   There may be azimuthal structure associated with Saturn, but since we recorded the data in the frame rotating with Neptune's mean motion, any such structure has been smeared into a ring.  We do not have great confidence in the details of this structure because Jupiter family comets may also contribute dust to this region of the cloud \citep{nesv09}, and we did not include this dust source in our model.

The submillimeter images tell a different story.   The submillimeter radiation comes mostly from the large grains ($s \gtrsim 100 \ \mu$m), which do not venture far from their initial conditions, even in the model with optical depth $10^{-7}$.  The submillimeter images mostly trace the distribution of source particles.  Figure~\ref{fig:submillimeter} shows that an observer looking at the KB dust  from a nearby star at submillimeter wavelengths would probably see a narrow ring of large grains at 42-45 AU.   This ring consists of large grains associated mostly with the cold classical Kuiper Belt.  
 
The azimuthally structures shown in the infrared images never become visible at submillimeter wavelengths, at any dust level.  Except for the increased numerical noise in the $\tau_{\rm{max}} \sim 10^{-4}$ simulations, the two images in Figure~\ref{fig:submillimeter}  are identical.  Indeed, Figure~\ref{fig:submillimeter} suggests that it would be impossible for an extra-solar observer to recognize evidence of Neptune in the KB dust if they only had submillimeter telescopes.

Though the $\tau_{\rm{max}} \sim 10^{-7}$ ring may resemble the classic ring + gap structure of \citet{lz99} and \citet{kuch03} in its geometric optical depth, the resemblance vanishes in the submillimeter.   The primary reason is that even when the disk has a total optical depth of $10^{-7}$, collisions destroy the grains probed by submillimeter imaging before they can venture far from their sources.   Collisions are important,  even in today's KB.

At no wavelength does the KB dust show the two-lobed ring characteristic of the plutino population (Figure~\ref{fig:collisionless}).  In our models, dust released by plutinos makes only a tiny contribution to the resonant structure in the Kuiper Belt cloud, primarily because the population of plutinos is a smaller fraction of the KB than previously realized \citep{hahn05}.

\section{Discussion}
\label{sec:discussion}

\subsection{Resonant Rings and Clumps}

Several authors have tried to understand exactly how important MMRs can be in sculpting the shapes of disks.   How clumpy can exozodiacal clouds be, and how will this clumpiness impact searches for extrasolar Earth-like planets \citep[e.g.][]{robe09}?   Do the clumps we see in millimeter and sumillimeter images of debris disks necessarily point to planets \citep[e.g.][]{wiln02}?  Our simulations shine some light on this problem.

For example, \citet{kriv07} divided resonant effects in debris disk dust into two categories: large grains released from source populations that are in MMRs, and small grains that become trapped in MMRs.   Our simulations reveal both of these effects.   \citet{kriv07} estimated that clumps created by small bodies in resonance would persist only at optical depths $\lesssim 10^{-4}$; at higher optical depths, they would be replaced by a narrow ring \citep[see also][]{plav09}.    Our simulations roughly confirm this prediction; we find that the clumps created by small dust grains in our simulation fade into azimuthally-symmetric rings at an optical depth of $\tau_r$ ($10^{-5}$ in the KB).

Our simulations also reveal a phenomenon relatively unanticipated by previous authors:  the importance of higher-order resonances ($n$:$n-2$, $n$:$n-3$, etc.) in sculpting a dusty disk.  We find that even at low dust levels, the first order resonances can become saturated, because, as \citet{star09} demonstrated, grain-grain collision rates are higher both in and near MMRs.   Also, the sources of the grains in our simulations are located near the planet in our simulations, where the MMRs are strong and dense; in many previous simulations, the grains were launched from two or three times the planet's semimajor axis.  Having the sources located near the planet as we do might promote trapping in higher order MMRs.


The higher-order resonances in the forest of MMRs near a planet are associated with many different geometries.  But all the geometries have a common feature:  they protect a particle from very close encounters with the planet.   The result is that when these resonances are populated, the ring they yield may extend inside and outside the planet's orbit, but  the location of the planet is always some kind of relatively dust-free gap. 


As we mentioned above, the actual dust level in the KB has never been directly measured.  But our models suggest one potentially easy way to measure the dust level:  take images of the KB dust, e.g., from a probe in the outer Solar System, and match them to the morphology of our models.  Use that data to search for a ring with the gap located at Neptune; its presence and strength would indicate the degree to which collisions remove small particles from resonances with Neptune.   Looking for an azimuthal asymmetry like this one might be easier than measuring the D.C. dust background.

\subsection{Limitations of the Simulations}
\label{sec:simulations}

We have done our best to emphasize interpretations of our models that we think will be robust.  However, our models represent only one step toward understanding the effect of collisions on the morphology of the Kuiper Belt dust cloud.  In the next two subsections, we will discuss some of the limitations of our simulations that should be kept in mind.   

Though our simulations cover three orders of magnitude in grain size, they neglect grains smaller than $1  \ \mu$m.   Most grains smaller than this size are ejected by radiation pressure in one dynamical timescale.  However, \citet{stru06} showed that populations of these so-called $\beta$-meteoroids can contribute substantially to the optical depth of a debris disk, especially the region exterior to the birth ring.  The absence of these small grains is probably especially important for images at short wavelengths; for this reason we decided to show images of our models at only far-IR and submillimeter wavelengths.

Some small grains ($\lesssim 1 \ \mu$m) in the Solar System originate in the interstellar medium, and fly rapidly through the Solar System on hyperbolic orbits.  Collisions between KBOs and these interstellar grains can be an important source of KB dust \citep{yama98}.  Our simulations do not explicitly model these high-velociy grains, which might also be important in destroying KB grains.

Our models also contain only a simplified treatment of particles larger than about $1400 \ \mu$m.   We chose this cutoff because debris disks generally become too faint to image long-ward of millimeter wavelengths.   Modeling large bodies in debris disks presents several complications.  Their PR lifetimes may be longer than the lifetime of the system.   Collisions with these bodies can produce long-lived fragments, or even enter the cratering regime.   We leave this work to future simulations.


A related issue is that our treatment of collisions contains no explicit treatment of fragmentation; it assumes that all grain production is associated with the source populations.  All the the explicitly modeled collisions between grains yield either complete vaporization or leave the particles unperturbed.   Though we estimate that typical collision velocities are high enough ($\sim 0.5$ km s${}^{-1}$) to make vaporization the most common collision outcome for small icy grains, there will be some collisions gentle enough, and some grains strong enough to lead to fragmentation.  Those fragments will often be $\beta$-meteoroids like those mentioned above.  Fragmentation will also be associated with some dissipation of orbital energy, a process we do not model.   

If fragmentation only serves to turn large bodies (i.e. source population bodies) into small ones (i.e ones that can be transported far from their source by radiation effects), then we are probably already modeling it accurately enough.  But to the degree that transportable bodies start fragmenting into other transportable bodies, we could be missing a key piece of physics.  One example of this situation is the creation of $\beta$-meteoroids, mentioned above.   Another possible situation that would demand a fragmentation model would be the case of low typical collision velocities, like a very dynamically cold disk.  


Finally, our models neglect Lorentz forces on dust grains, and drag from any interstellar gas.   These forces can potentially be important, particularly for charged grains, \citep[e.g.][]{holm03} and for grains in the very outskirts of debris disks \citep[e.g.][]{debe09}.   We chose to neglect them because we felt that adding too many new ingredients at once to these complex simulations might make our results too hard to for us to physically interpret.    Moreover, the parameters of these effects range widely among various stars, so including them might hinder the use of our model as a baseline for comparison with other systems.

\subsection{Input Parameters and Interpretation}
\label{sec:input}

Our models of the Kuiper Belt, scaled up in mass, move beyond simple linear scalings, helping us compare the Kuiper Belt to extrasolar debris disks.   But this comparison remains far from perfect.   Other debris disks, for example, are not the same age as the Kuiper belt; they could represent younger systems populated by source bodies doomed to vanish via collisions and orbital instabilities.   Or they could contain planets in the process of migration; this process could influence the morphology of a debris cloud \citep{wyat03, murr05}.  A few debris disks might even represent a transient event, perhaps analogous to late heavy bombardment in the Solar System \citep{boot09}.  

Another important point is that the source distributions we chose might not be a good representation of the actual distribution of dust sources in the Kuiper belt.   First of all, the distribution of KBOs is not completely known.   For example, there is a strong bias toward detecting KBOs with small perihelion distances, so we do not yet have a good inventory of dynamically-cold KBOs beyond about 46 AU.    Since, according to our simulations, the dynamically cold KBOs contribute most of the particles trapped in MMRs, this lack of knowledge hampers our ability to predict the population of dust in the 2:1 MMR with Neptune ($\sim 48$ AU), and any MMRs exterior to that.  

But even if we knew the exact orbital distribution of KBOs, down to, say 1 km in size, this distribution would not correspond exactly to where the dust production events occur.   Some dust production in the KB occurs when ISM grains hit KBOs \citep{yama98}; our source populations probably represent this mechanism well.   But if the KB is like the asteroid belt, then some dust production is probably associated with collisional families, perhaps like the Haumea family \citep{brow07}.  Therefore, for the time being, it seems appropriate that we content ourselves with simple parametric source distributions, inspired by KBO observations, like those we used here.

\section{Conclusions}
\label{sec:conclusions}

We modeled the 3-D dust distribution in the Kuiper Belt taking into account perturbations from Jupiter, Saturn, Uranus and Neptune, the destruction of dust grains via collisions, and the interaction of these phenomena, including the enhanced destruction of grains in mean motion resonances.   We demonstrated that the collisional grooming algorithm can approximately reproduce the \citet{dohn69} collisional equilibrium size distribution---though resonant trapping tends to modify the size distribution in a resonant ring.  The dust level in the KB has never been directly measured;  we suggested that one way to measure the dust level would be matching images of the KB dust, e.g., from a probe in the outer Solar System, to the morphology of our models.  Searching for a ring with the gap at Neptune in this manner might be easier than measuring the D.C. dust background.

Here are the primary conclusions we have drawn from our models, about the Kuiper Belt dust population itself and about debris disks in general.

\begin{itemize}
 

\item An observer looking at the KB dust from a nearby star at submillimeter wavelengths (Figure~\ref{fig:submillimeter}) would probably see an azimuthally-symmetric ring of large grains at 42-45 AU.   This ring consists of large grains,  $> 100 \  \mu$m in size, associated with the cold classical Kuiper Belt.     The submillimeter morphology of the KB is largely independent of optical depth, because collisions limit the lifetimes of large particles, even at optical depths of $10^{-7}$.   

\item An observer looking at the KB dust if it had optical depth $10^{-4}$ in the far infrared would see roughly the same ring as in the submillimeter images; an azimuthally-symmetric ring at 40--47 AU (Figure~\ref{fig:ir}).  

 \item  At lower optical depths, $\tau_{\rm{max}} \sim 10^{-6}$ or $\tau_{\rm{max}} \sim 10^{-7}$, the ring seen in the infrared widens and a gap opens at the location of Neptune.  This ring consists of dust trapped in MMRs with Neptune, and corresponds to the ring predicted by \citet{lz99}.  A secondary ring appears as well near Saturn's orbit.   Some grains are trapped in the 2:1, 3:2, and other MMRs with Saturn, but mostly this secondary ring appears because Saturn clears a central hole in a centrally-illuminated disk.   This secondary central ring of small dust grains may be analogous to the hot central dust cloud seen around Fomalhaut \citep{stap04, absi09}.

  

\item Mean motion resonances can contribute strongly to the appearance of debris disks, despite previous suggestions to the contrary.  They can contribute in two ways:  1) large bodies that dominate the submillimeter images remain near their source, which may itself be resonant, like the Plutinos.  2)  smaller grains become trapped in MMRs as they spiral into the star.   Though the small-grain dust population of the first-order resonances can saturate in the presence of collisions, these smaller particles also interact with a forest of higher-order MMRs.  The higher-order MMRs also serve to create a ring-like structure near the orbit of the perturbing planet.   

\item  At high optical depths, debris disk images are likely dominated by the birth ring, the source of the particles.   But at optical depths below $\tau_r \sim v/c$, small dust grains trapped in mean motion resonances can dominate the images.   Here $v$ is the local Keplerian speed, and $c$ is the speed of light.   Hot dust (exozodiacal dust) in the centers of cold debris disks will never exceed this crossover optical depth unless there is a new source of particles near the star.

\item Though we can see its signature in the semimajor axis distributions of the particles, we find that dust released by plutinos makes a negligible contribution to images Kuiper Belt cloud, primarily because the population of plutinos is a smaller fraction of the dust cloud than previously realized \citep{hahn05}.   


\item Grain-grain collisions are important in setting the 3-D shape of {\it today's} Kuiper Belt dust cloud, even at an optical depth of $\sim 10^{-7}$.  
 
\end{itemize}

Our simulations show us how our Solar System might appear to an extra-solar observer, depending on how much dust there actually is in the Kuiper Belt.   They illustrate how our changing picture of the KBO orbital distribution has changed our picture of the KB dust.   However, our models leave many potentially important questions unanswered.  How do $\beta$ meteoroids, comets, interstellar grains, and fragmentation of small grains affect the appearance of the KB disk?  How do collisional families and other transient phenomena influence the KB dust?  

The biggest outstanding issue for this kind of model is probably the question of how and whether resonant populations of source bodies create dust clumps.   For example, while, as we showed, increased collision rates among grains from the plutinos serves to reduce the azimuthal asummetry in cloud, erasing clumps.   But at the same time, when the overall mass of the disk is increased, the collision rate among the plutinos themselves, increases as well, possibly increasing the azimuthal asymmetry of the cloud.   This phenomenon may explain the presence of the clumpy rings seen in highly collisional debris disks like $\epsilon$ Eridani.   No models can yet calculate which effect dominates, but perhaps the next generation of models, together with more observations of the Kuiper Belt and other debris disks, will help answer these questions.

Two observational goals stand out that would especially help us take the next steps with our models:  1) we would like to know the population of KBOs in the 2:1 MMR with Neptune at low eccentricity, and 2) we would like to have more resolved images of debris disks with low optical depth ($\lesssim 10^{-5}$) where $\tau < \tau_r$.  We expect that survey telescopes like the Large Synoptic Survey Telescope (LSST) may help achieve the first goal, and that deep thermal and coronagraphic imaging of debris disks, e.g., with the Herschel space telescope, the Atacama Large Millimeter Array (ALMA) and the James Webb Space Telescope (JWST), should help attain the second.

\acknowledgments

We thank J.J. Kavelaars and Mike Brown for helpful discussions about the KBO orbital distribution, and Alexander Krivov for encouragement.   We thank the NASA High-End Computing Program for granting us time on the Discover cluster.   Christopher Stark was supported in part by the NASA GSRP Program at Goddard Space Flight Center.

\newpage

\end{document}